\documentclass[10pt,nofootinbib,aps,prd]{revtex4-1}
\usepackage{amsmath,mathptmx,amssymb}
\usepackage{CJK}
\usepackage{graphicx,graphics,subfigure}
\usepackage{tabularx,ctable}
\usepackage[colorlinks=true,linkcolor=blue,citecolor=blue,urlcolor=blue]{hyperref}

\DeclareSymbolFont{symbols} {OMS}{cmsy}{m}{n}

\def\be{\begin{equation}}
\def\ee{\end{equation}}
\def\bea{\begin{eqnarray}}
\def\eea{\end{eqnarray}}
\def\ba{\begin{aligned}}
\def\ea{\end{aligned}}
\def\nn{\nonumber}
\def\p{\partial}

\begin{document}


\title{Thermodynamic topological classes of the rotating, accelerating black holes}

\author{Wentao Liu$^{1}$}

\author{Li Zhang$^{1}$}

\author{Di Wu$^{2}$}
\email{Corresponding author: wdcwnu@163.com}

\author{Jieci Wang$^{1}$}
\email{Corresponding author: jcwang@hunnu.edu.cn}

\affiliation{$^{1}$Department of Physics, Key Laboratory of Low Dimensional Quantum Structures and Quantum Control of Ministry of Education, and Synergetic Innovation Center for Quantum Effects and Applications, Hunan Normal University, Changsha, Hunan 410081, People's Republic of China \\
$^{2}$School of Physics and Astronomy, China West Normal University, Nanchong, Sichuan 637002, People's Republic of China}

\date{\today}

\begin{abstract}
In this paper, we investigate the topological numbers for the rotating, accelerating neutral black hole and its AdS extension, as well as the rotating, accelerating charged black hole and its AdS extension. We find that the topological number of an asymptotically flat accelerating black hole consistently differs by one from that of its non-accelerating counterpart. Furthermore, we show that for an asymptotically AdS accelerating black hole, the topological number is reduced by one compared to its non-accelerating AdS counterpart. In addition, we demonstrate that within the framework of general relativity, the acceleration parameter and the negative cosmological constant each independently add one to the topological number. However, when both factors are present, their effects neutralize each other, resulting in no overall change to the topological number.
\end{abstract}

\maketitle

\section{Introduction}

The study of the properties of black hole spacetimes and gravitational fields fundamentally relies on exact solutions to the gravitational field equations.
Among the well-known family of exact black hole solutions to the four-dimensional Einstein field equations in General Relativity (GR), the most general Petrov type-D electrovacuum black hole solution is the Plebanski-Demianski metric \cite{AP98-98}.
The simplest case of this metric, non-rotating, uncharged, and without NUT charge or cosmological constant, is referred to as the C-metric \cite{PRD2-1359,IJMPD15-335}.
This class of solutions is also known as the accelerating black holes, which were interpreted by Kinnersley and Walker \cite{PRD2-1359} as the gravitational field produced by a point mass undergoing uniform acceleration, i.e., a pair of black holes accelerating away from each other in opposite directions \cite{GRG15-535}.
In recent years, a wide range of studies has focused on the different features of accelerating black holes, including their global causal structure \cite{CQG23-6745}, quasinormal modes \cite{PRD109-084049,JHEP1022047,JHEP0224140,
JHEP0224189,2305.04040,2408.05957}, quantum thermal properties \cite{PLA209-6}, holographic heat engines \cite{EPJC78-645,JHEP0219144}, weak/strong cosmic censorship conjecture \cite{PLB849-138433,SCPMA66-280412}, black hole shadows \cite{PRD103-025005}, and holographic complexity \cite{PLB823-136731,PLB838-137691}, etc \cite{PRL126-111601,JHEP0122102,PRL130-091603,JHEP1022074,JHEP0324050,JHEP0324079,2404.04691,
1711.04522,1906.09478,2403.09757,1408.3124,2205.15777,2211.06517,2305.03765,2309.13656,2411.12513}.
Especially, in Refs. \cite{PRL117-131303,JHEP0517116,PRD98-104038,JHEP0419096,PLB796-191,CQG38-145031,
CQG38-195024,PRD104-086005,2306.16187,2407.21329,2406.01897}, the thermodynamics of four-dimensional accelerating (charged and rotating) AdS black holes was thoroughly studied, leading to extensions of the first law of thermodynamics \cite{PRD7-2333,PRD13-191}, the Bekenstein-Smarr mass formula \cite{PRL30-71}, and the Christodoulou-Ruffini-like squared-mass formula \cite{PRL25-1596, PRD4-3552,PRD101-024057,PRD102-044007,PRD103-044014,JHEP1121031} to such spacetimes.\footnote{Recently, these formulas were perfectly extended to the Lorentzian Taub-NUT spacetimes \cite{PRD100-101501,PRD105-124013,PRD108-064034,PRD108-064035,PLB846-138227}.}

The development of the above mass formulas represents just one facet of black hole thermodynamics.
Quite recently, topology has emerged as a powerful mathematical tool, attracting substantial interest for its role in analyzing the thermodynamic phase transitions of black holes \cite{PRD105-104003,PRD105-104053,PLB835-137591,PRD107-046013,PRD107-106009,JHEP0623115,
2305.05595,2305.05916,2305.15910,PRD106-064059,PRD107-044026,PRD107-064015,2212.04341,
2302.06201,2304.14988,2309.00224,2312.12784,2402.18791,2404.02526,2407.09122,2407.20016}.
\footnote{Topology has also proven effective in the study of the properties of light rings \cite{PRL119-251102,PRL124-181101,PRD102-064039,PRD103-104031,PRD104-044019,PRD105-024049,
PRD105-064070,PRD108-104041,PRD109-064050,2408.05569,2412.18083} and timelike circular orbits \cite{PRD107-064006,JCAP0723049,2406.13270}.}
In Ref. \cite{PRL129-191101}, Wei \textit{et al.} proposed a novel and systematic classification scheme for black holes by considering them as thermodynamic topological defects, categorized according to their respective topological numbers in a pioneering manner.
A brief account of this method is presented below.

Hawking temperature $T$ and Bekenstein-Hawking entropy $S$ are two very important quantities in black hole thermodynamics.
In general, the standard Holmholtz free energy $F$ can be expressed as $F = M -TS$ \cite{PRD15-2752,PRD33-2092,PRD105-084030,PRD106-106015}.
Inspired by this, we introduce the generalized off-shell Helmholtz free energy, which can be expressed in the following form \cite{PRL129-191101}:
\be\label{FE}
\mathcal{F} = M -\frac{S}{\tau} \, ,
\ee
where the extra variable $\tau$ can be interpreted as the inverse temperature of the cavity surrounding the black hole. \footnote{\textcolor{black}{In a recent preprint \cite{2504.00503}, the authors discovered two special cases of the generalized Helmholtz free energy, corresponding to the generalized Helmholtz free energy in bumblebee gravity \cite{PRD39-683} and three-dimensional new massive gravity \cite{PRL102-201301,RMP84-671}. Therefore, when addressing the thermodynamic topology classification of black holes in these two theories, the form of the generalized free energy may require modification.}}
The generalized Helmholtz free energy only exhibits on-shell characteristics and returns to the standard Helmholtz free energy of the black hole when $\tau = T^{-1}$.
According to Ref. \cite{PRL129-191101}, one can construct a two-dimensional vector field $\phi = (\phi^r, \phi^\Theta)$ from the generalized Helmholtz free energy $\mathcal{F}$, where
\bea\label{vector}
\phi = \Big(\frac{\p \mathcal{F}}{\p r_h}\, , ~ -\cot\Theta\csc\Theta\Big) \, ,
\eea
in which $r_h$ is the radius of the black hole's event horizon, and $\Theta$ is an additional factor, with $\Theta \in [0,+\infty]$.
It is important to note that the component $\phi^\Theta$ becomes infinite at $\Theta = 0$ and $\Theta = \pi$, indicating that the vector points outward in both cases.

\textcolor{black}{Using Duan's $\phi$-mapping topological current theory \cite{SS9-1072,NPB514-705,PRD61-045004},
a topological current can be described as follows \footnote{\textcolor{black}{Unlike topological methods based on light rings \cite{PRL119-251102,PRL124-181101} or timelike circular orbits \cite{PRD107-064006}, which prioritize dynamical or geometric properties, the $\phi$-mapping theory offers distinct advantages in probing thermodynamic features. Specifically, our work employs $\phi$-mapping theory to classify black hole solutions by analyzing the global topology of thermodynamic defects, establishing a unified framework for black hole solution classification \cite{PRL129-191101}. This contrasts with other applications of $\phi$-mapping theory that focus on characterizing critical points during black hole phase transitions \cite{PRD105-104003}. While these alternative approaches excel at dissecting phase behavior, our work emphasizes the intrinsic classification of black hole solutions through their thermodynamic defect structures, thereby complementing existing analyses of dynamic or geometric traits.}}:
\be\label{jmu}
j^{\mu}=\frac{1}{2\pi}\epsilon^{\mu\nu\rho}\epsilon_{ab}\p_{\nu}n^{a}\p_{\rho}n^{b}\, , \qquad
\mu,\nu,\rho=0,1,2,
\ee
where $\p_{\nu}= \p/\p{}x^{\nu}$ and $x^{\nu}=(\tau,~r_h,~\Theta)$. The unit vector $n$ formulates as $n = (n^r, n^\Theta)$, where $n^r = \phi^{r_h}/||\phi||$ and
$n^\Theta = \phi^{\Theta}/||\phi||$. It is easy to indicate that the topological current (\ref{jmu}) given above is conserved,
allowing one to easily deduce $\p_{\mu}j^{\mu} = 0$. It is then demonstrate that the topological
current $j^\mu$ is a $\delta$-function of the field configuration \cite{NPB514-705,PRD61-045004}:
\be
j^{\mu}=\delta^{2}(\phi)J^{\mu}\Big(\frac{\phi}{x}\Big)\, ,
\ee
where the 3-dimensional Jacobian $J^{\mu}\big(\phi/x\big)$ is defined as: $\epsilon^{ab}J^{\mu}
\big(\phi/x\big) = \epsilon^{\mu\nu\rho}\p_{\nu}\phi^a\p_{\rho}\phi^b$. It is easy to obtain
that $j^\mu$ equals to zero only when $\phi^a(x_i) = 0$.} Therefore, for a given parameter region $\rho$, the topological number is expressed as
\be
W = \int_{\rho}j^{0}d^2x = \sum_{i=1}^{N}\beta_{i}\eta_{i} = \sum_{i=1}^{N}w_{i}\, ,
\ee
where $w_i$ represents the winding number at the $i$th point of $\phi$, $j^{0}$ denotes the density of the topological current, $\beta_i$ is the positive Hopf index, and the Brouwer degree $\eta_{i}= \mathrm{sign}(J^{0}({\phi}/{x})_{z_i})=\pm 1$, respectively.
Note that the local winding number $w_i$ is regarded as a crucial tool for determining local thermodynamic stability.
\textcolor{black}{The positive or negative heat capacity of a black hole state can be shown to, respectively, correspond to a positive or negative winding number, $w = +1$ and $w = -1$, associated with locally stable and unstable black hole states.}
The global topological number $W$ represents the difference between the counts of thermodynamically stable and unstable black holes at a specific temperature.
\textcolor{black}{Therefore, the global topological number $W$ characterizes each black hole system and can be used as a classification parameter.}

The elegance and general applicability of this method led to its widespread adoption, facilitating the exploration of topological numbers across different black hole models \cite{PRD107-064023,JHEP0123102,PRD107-024024,PRD107-084002,PRD107-084053,2303.06814,
2304.05695,2306.05692,2306.11212,EPJC83-365,2306.02324,
AP458-169486,2310.09602,2310.15182,2311.04050,2311.11606,
2312.04325,2312.06324,2312.13577,2312.12814,PS99-025003,2402.00106,PLB856-138919,
AP463-169617,PDU44-101437,2403.14167,PDU46-101617,2405.02328,2405.07525,2405.20022,2406.08793,
2407.05325,2408.08325,2409.04997}.
However, all previous research has primarily focused on non-accelerating cases, overlooking the topological numbers of accelerating black holes.
Very recently, we have extended the topological approach to include non-rotating, accelerating black holes, considering both neutral and charged cases, as well as their AdS extensions \cite{PRD108-084041}.
\textcolor{black}{Extending this work to more general rotating, accelerating black
holes will clarify how the acceleration parameter influences their topological numbers.} This also serves as the motivation for the current study.

In this paper, we explore, within the framework of the topological method for black hole thermodynamics, the topological numbers of the rotating, accelerating neutral black hole and its AdS counterpart, as well as the rotating, accelerating charged black hole and its AdS extension, via the generalized off-shell Helmholtz free energy.
We found that, for the thermodynamic topological numbers of black hole solutions within the framework of general relativity, while both the acceleration parameter and the negative cosmological constant independently increase the topological number by one, their effects counterbalance each other when both are present in a black hole solution, which leads us to conjure that it might also hold true for the accelerating AdS black holes in alternative theories of gravity.

\textcolor{black}{Our analysis now emphasizes that prior studies focused on non-accelerating black holes or static accelerating cases, while this work extends the topological classification to rotating, accelerating black holes in both asymptotically flat and AdS spacetimes. We also highlight the discovery that the acceleration parameter and cosmological constant counterbalance each other's topological contributions--a novel result not explored in earlier work.}

The remaining part of this paper is organized as follows.
In section \ref{II}, we briefly introduce the rotating, accelerating charged AdS black hole.
Section \ref{III} begins with an exploration of the topological number of the rotating, accelerating black hole, focusing on the Kerr C-metric solution, and then extends this analysis to the Kerr-AdS C-metric solution, where a negative cosmological constant is present.
In section \ref{IV}, we examine the topological number of the rotating, accelerating charged black hole through the Kerr-Newman C-metric (KN C-metric) solution and subsequently extend this investigation to the KN-AdS C-metric case.
Finally, section \ref{V} presents our conclusions and future outlooks.

\section{A brief introduction of rotating charged AdS accelerating black hole}\label{II}

\begin{figure}[t]
\centering
\includegraphics[width=0.6\textwidth]{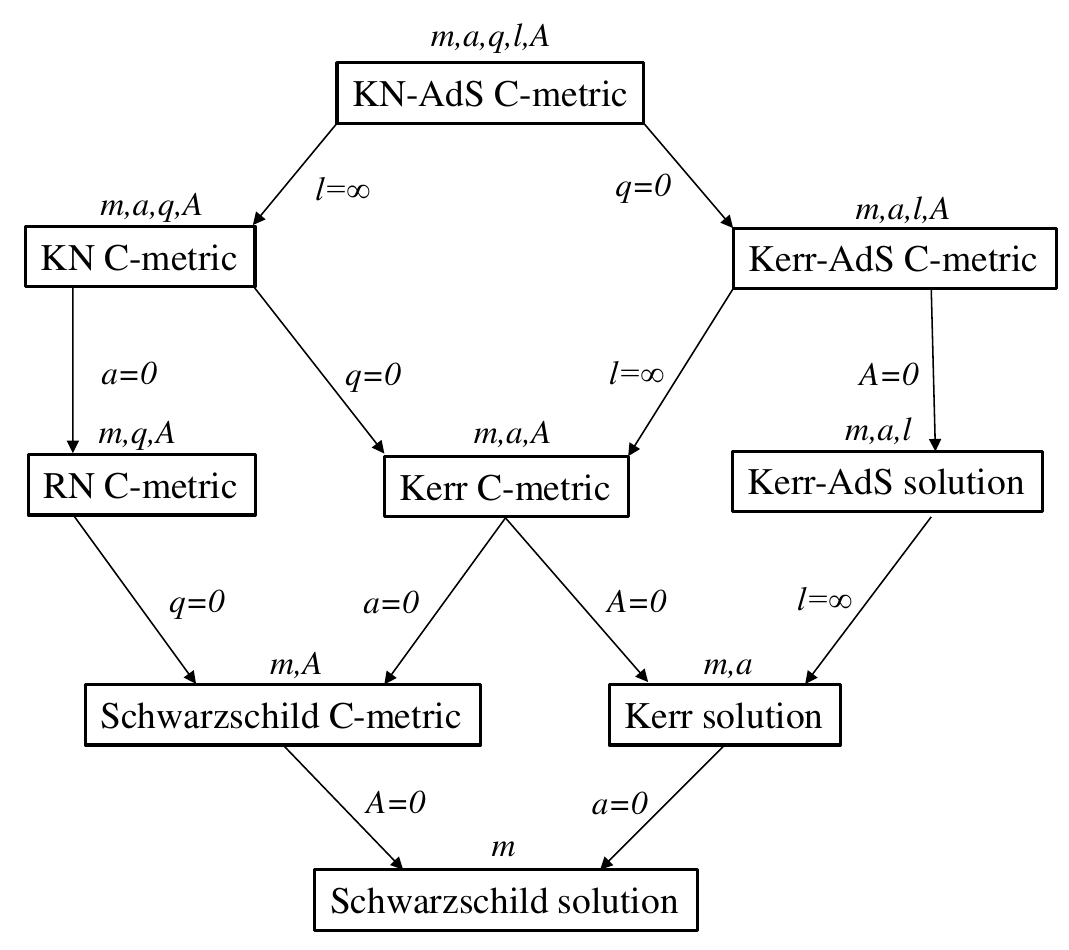}
\caption{The structure of the family of black hole solutions represented by metric (\ref{KNAdSCmetric}). This family has five parameters $m$, $a$, $q$, $l$, $A$.
\label{fig1}}
\end{figure}

We start by introducing the generalized AdS C-metric solution, derived from the Plebanski-Demianski metric \cite{AP98-98} in Boyer-Lindquist type coordinates \cite{IJMPD15-335}, which includes rotation, electric charge, cosmological constant $\Lambda = -3/l^2$, and the corresponding Abelian gauge potential, expressed as \cite{JHEP0419096}
\bea
ds^2 &=& \frac{1}{H^2}\Bigg\{-\frac{f(r)}{\Sigma}\left(\frac{dt}{\alpha} -a\sin^2\theta \frac{d\varphi}{K}\right)^2 +\frac{\Sigma}{f(r)}dr^2 +\frac{\Sigma }{h(\theta)}d\theta^2 \nn \\
&&+\frac{h(\theta)\sin^2\theta}{\Sigma}\left[\frac{adt}{\alpha} -(r^2 +a^2)\frac{d\varphi}{K} \right]^2 \Bigg\} \, , \label{KNAdSCmetric} \\
F &=& dB \, , \qquad B = -\frac{qr}{\Sigma}\left(\frac{dt}{\alpha} -a\sin^2\theta\frac{d\varphi}{K} \right) +\frac{qr}{(r^2 +a^2)\alpha}dt \, , \label{Abelian}
\eea
where $\alpha$ is a rescaled factor, and the metric functions are
\bea\label{mf}
&&f(r) = (1 -A^2r^2)(r^2 -2mr +a^2 +q^2) +\frac{r^4 +a^2r^2}{l^2} \, , \nn  \\
&&h(\theta) = 1 +2mA\cos\theta +\left[A^2(a^2 +q^2) +\frac{a^2}{l^2} \right]\cos^2\theta \, ,  \\
&&\Sigma = r^2 +a^2\cos^2\theta \, , \qquad H = 1 +Ar\cos\theta \, . \nn
\eea
in which $K$ is the conical deficit of the spacetime, $m$, $a$, $q$, $A$, $l$ are the mass, rotation, electric charge, acceleration parameters of the black hole and the AdS radius, respectively.
The event horizon is located at the largest root of equation: $f(r_h) = 0$.
For the special case in which $a$, $q$, $A$ vanish and $l\to\infty$, the general structure of this family of black hole solutions is given in figure \ref{fig1}.
The $K$ factor is typically integrated into the azimuthal coordinate, resulting in an arbitrary periodicity, typically determined by a regularity condition at one of the poles.
However, here the $K$ factor is explicitly included to ensure that the periodicity of $\varphi$ remains fixed at $2\pi$.
The conical deficit at each axis is then determined by analyzing the behavior of the $\theta$-$\varphi$ portion of the metric near the south pole $\theta = \theta_+ = 0$ and the north pole $\theta = \theta_- = \pi$, respectively \cite{JHEP0419096}:
\be
ds_{\theta,\varphi}^2 \propto d\theta^2 +h^2(\theta)\sin^2\theta\frac{d\varphi^2}{K^2} \sim d\vartheta^2 + (\Xi\pm 2mA)^2\vartheta^2d\varphi^2\, ,
\ee
where $\vartheta_\pm = \pm(\theta-\theta_\pm)$ provides a local radial coordinate near each axis, and
\be
\Xi = 1 - a^2/l^2 + A^2(q^2+a^2)\, .
\ee
The above geometry indicates that there exists a conical singularity at the symmetry axis, with the tensions of the cosmic strings being
\be
\mu_{\pm} = \frac{1}{4}\left(1 -\frac{\Xi \pm 2mA}{K} \right) \, .
\ee
Thus, it is easy to see that the acceleration arises from the difference in conical deficits between the north and south poles:
\be
\mu_- -\mu_+ = mA/K \, ,
\ee
where $K$ represents an overall deficit in the spacetime:
\be
\bar{\mu} = \frac{1}{2}(\mu_+ +\mu_-) = \frac{1}{4} -\frac{\Xi}{4K} \, .
\ee

\section{Topological classes of rotating neutral accelerating black hole}\label{III}
In this section, we will discuss the topological number of the four-dimensional rotating, accelerating neutral black hole by considering the Kerr C-metric black hole, and then extend the analysis to the case of the Kerr-AdS C-metric black hole with a nonzero negative cosmological constant.

\subsection{Kerr C-metric black hole}\label{3.1}
The metric of the Kerr C-metric black hole is still given by Eq. (\ref{KNAdSCmetric}), and now the metric functions $f(r)$ and $h(\theta)$ reduce to $f(r) = (1 -A^2r^2)(r^2 -2mr +a^2)$ and $h(\theta) = 1 +2mA\cos\theta +A^2a^2\cos^2\theta$, respectively.
The thermodynamic quantities are \cite{JHEP0419096}
\bea\label{ThermKerrC}
&&M = \frac{m(1 -A^2a^2)}{\alpha K(1 +A^2a^2)} \, , ~ \qquad T = \frac{\p_{r_h}f(r_h)}{4\pi\alpha(r_h^2 +a^2)} \, , ~ \qquad S = \frac{\pi(r_h^2 +a^2)}{K(1 -A^2r_h^2)} \, ,   \nn \\
&&\alpha = \frac{\sqrt{1 -A^2a^2}}{\sqrt{1 +A^2a^2}} \, , ~ \qquad J = \frac{ma}{K^2} \, , ~ \qquad \mu_{\pm} = \frac{1}{4}\left[1 -\frac{1\pm 2mA +A^2a^2}{K} \right] \, ,  \\
&&\Omega = \frac{aK}{\alpha(r_h^2 +a^2)} -\frac{A^2aK}{\alpha(1 +A^2a^2)} \, , \quad \lambda_\pm = \frac{r_h}{\alpha(1\pm Ar_h)} -\frac{MK}{1 +A^2a^2} \mp\frac{Aa^2}{\alpha(1 +A^2a^2)} \, . \quad \nn
\eea
\textcolor{black}{Here and hereafter (in Eqs. (\ref{ThermKerrAdSC}), (\ref{ThermKNC}), and (\ref{ThermKNAdSC})), $M$, $S$, $T$, $\Omega$, and $J$ represent the mass, Bekenstein-Hawking entropy, Hawking temperature, angular velocity, and angular momentum, respectively; $\mu_\pm$ are the tensions of the conical deficits on the north and south poles, and $\lambda_\pm$ are the thermodynamic lengths conjugate to the tensions $\mu_\pm$.}
It is a simple matter to check that the above thermodynamic quantities simultaneously satisfy the first law and the Bekenstein-Smarr mass formula
\bea
dM &=& TdS +\Omega dJ -\lambda_+d\mu_+ -\lambda_-d\mu_- \, , \\
M &=& 2TS +2\Omega\, J \, .
\eea

In the following, we will derive the topological number of the Kerr C-metric black hole.
The generalized off-shell Helmholtz free energy is
\be
\mathcal{F} = \frac{(r_h^2 +a^2)(1 -A^2a^2)}{2\alpha Kr_h(1 +A^2a^2)} -\frac{\pi(r_h^2 +a^2)}{K\tau(1 -A^2r_h^2)} \, ,
\ee
Using the definition of Eq. (\ref{vector}), the components of the vector $\phi$ can be easily computed as follows:
\bea
\phi^{r_h} &=& \frac{(r_h^2 -a^2)(1 -A^2a^2)}{2\alpha Kr_h^2(1 +A^2a^2)} -\frac{2\pi r_h(1 +A^2a^2)}{K\tau(A^2r_h^2 -1)^2} \, , \\
\phi^{\Theta} &=& -\cot\Theta\csc\Theta \, .
\eea
By solving the equation: $\phi^{r_h} = 0$, one can obtain the zero point of the vector field $\phi^{r_h}$ as
\be\label{tauKerrC}
\tau = \frac{4\pi\alpha r_h^3(1 +A^2a^2)^2}{(r_h^2 -a^2)(1 -A^2a^2)(A^2r_h^2 -1)^2} \, .
\ee
It is important to note that Eq. (\ref{tauKerrC}) consistently reduces to the result obtained in the case of the four-dimensional Kerr black hole \cite{PRD107-024024} when the acceleration parameter $A$ vanishes.

\begin{figure}[t]
\subfigure[]
{\label{fig2}
\includegraphics[width=0.4\textwidth]{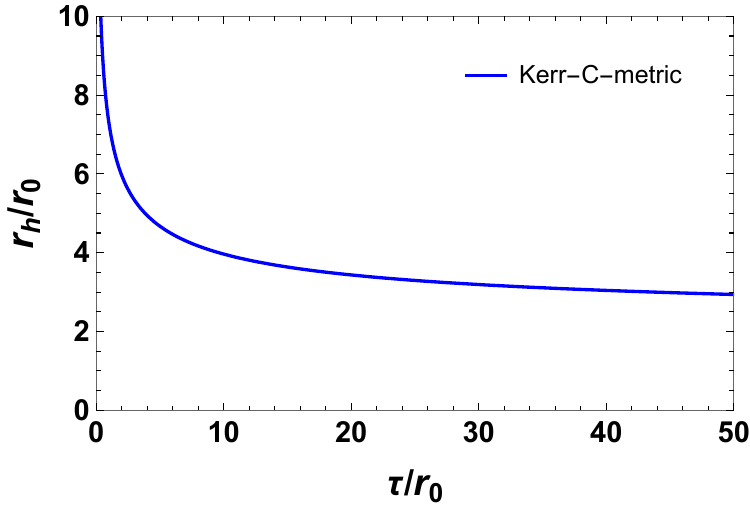}}
\subfigure[]
{\label{fig3}
\includegraphics[width=0.4\textwidth]{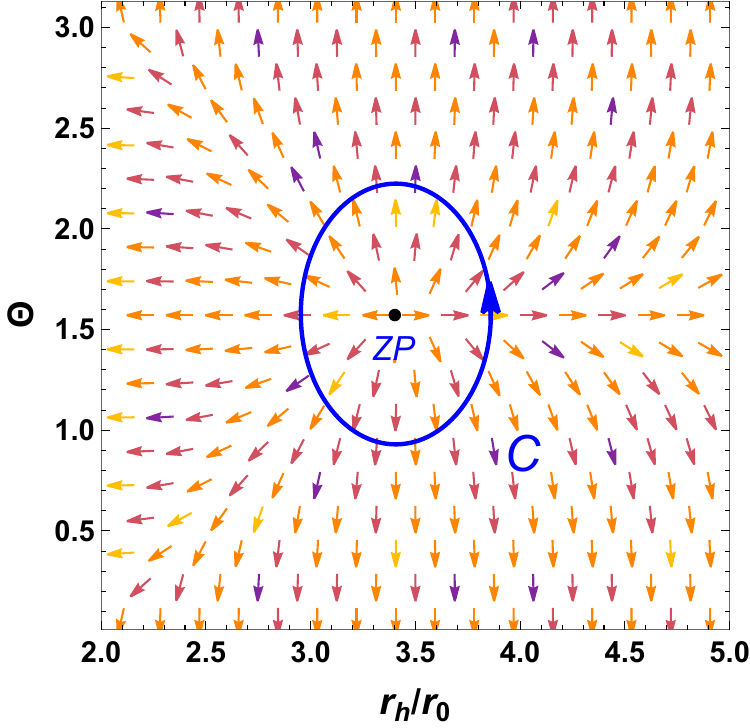}}
\caption{(a) Zero points of the vector $\phi^{r_h}$ shown in the $r_h-\tau$ plane with $a/r_0 = 1$ and $Ar_0 = 0.5$. \textcolor{black}{There is always one thermodynamically stable four-dimensional Kerr C-metric black hole for any value of $\tau$. Obviously, the topological number is: $W = 1$; (b) The arrows represent the unit vector field $n$ on a portion of the $r_h-\Theta$ plane for the four-dimensional Kerr C-metric black hole with $a/r_0 = 1$, $Ar_0 = 0.5$, and $\tau/r_0 = 20$.} The zero point (ZP) marked with black dot is at $(r_h/r_0, \Theta) = (3.44,\pi/2)$. The blue contours $C$ is closed loop enclosing the ZP.}
\end{figure}

Assuming $Ar_0 = 0.5$, $a/r_0 = 1$ for the Kerr C-metric black hole, we plot Figs. \ref{fig2} and \ref{fig3} to illustrate key aspects of the system.
These figures display the zero points of the component $\phi^{r_h}$ and the behavior of the unit vector field $n$ on a segment of the $\Theta-r_h$ plane, with $\tau = 20r_0$, where $r_0$ is an arbitrary length scale defined by the size of the cavity enclosing the rotating, accelerating black hole.

From Fig. \ref{fig2}, it is apparent that for any given value of $\tau$, there is only one thermodynamically stable Kerr C-metric black hole.
This sharply contrasts with the four-dimensional Kerr black hole, which exhibits two distinct thermodynamic behaviors: one black hole branch that is stable and another that is unstable \cite{PRD107-024024}.
This distinction underscores the significant influence of the acceleration parameter on the thermodynamical numbers of the rotating neutral black holes.

In Fig. \ref{fig3}, the zero point is found at $(r_h/r_0, \Theta) = (3.44,\pi/2)$.
As a result, the winding number $w_i$ for the blue contour $C$ is identified as $w_1 = 1$, which differ from those associated with the four-dimensional Kerr black hole \cite{PRD107-024024}.
Regarding the global topological characteristics, the topological number $W = 1$ for the Kerr C-metric black hole can be directly seen in Fig. \ref{fig3}, distinguishing it from the topological number of the Kerr black hole ($W = 0$).
Therefore, it can be inferred that the Kerr C-metric black hole and the Kerr black hole differ not only in geometric topology but also belong to distinct categories in thermodynamic topology.
\textcolor{black}{The underlying mechanism responsible for these distinct properties may be attributed to the acceleration parameter, which introduces additional geometric effects via conical deficits. This modifies the root structure of $f(r)$, consequently reducing the number of unstable branches. This physical intuition suggests that acceleration could suppress the complexity of thermodynamic phase transitions by stretching spacetime along the symmetry axis.}
Consequently, it would be compelling to investigate the topological characteristics of black holes with unusual geometries, such as the type-D NUT C-metric black hole \cite{2409.06733}, though this first requires establishing its consistent thermodynamics.
In addition, combining the intriguing conclusion presented in Ref. \cite{PRD108-084041}, which states that the difference in topological numbers between an asymptotically flat, static, accelerating black hole and its corresponding non-accelerating counterpart is always unity, we suggest a more general conjecture within the framework of general relativity.
Specifically, we propose that the difference in topological numbers between any asymptotically flat accelerating black hole and its corresponding non-accelerating black hole is consistently unity.
This inference will be further substantiated in section \ref{4.1}, where we compute the topological number of the KN C-metric black hole and compare it with that of the KN black hole, thereby demonstrating the validity and applicability of this conjecture.

\subsection{Kerr-AdS C-metric black hole}\label{3.2}
In this subsection, we will extend the above discussions to the cases of the rotating, accelerating neutral AdS black hole by considering the Kerr-AdS C-metric black hole,
whose metric is still given by Eq. (\ref{KNAdSCmetric}), but now
\be\ba
f(r) &= (1 -A^2r^2)(r^2 -2mr +a^2) +\frac{r^4 +a^2r^2}{l^2} \, ,  \\
h(\theta) &= 1 +2mA\cos\theta +\left[A^2a^2 +\frac{a^2}{l^2} \right]\cos^2\theta \, ,
\ea\ee
where the AdS radius $l$ is associated with the thermodynamic pressure $P = 3/(8\pi l^2)$ of the four-dimensional AdS black holes \cite{CPL23-1096,CQG26-195011,
PRD84-024037}.

The thermodynamic quantities are \cite{JHEP0419096}
\be\ba\label{ThermKerrAdSC}
&M = \frac{m(\Xi +a^2/l^2)(1 -A^2l^2\Xi)}{\Xi\alpha K(1 +A^2a^2)} \, , \quad T = \frac{\p_{r_h}f(r_h)}{4\pi\alpha(r_h^2 +a^2)} \, , ~ \quad S = \frac{\pi(r_h^2 +a^2)}{K(1 -A^2r_h^2)} \, ,   \\
&J = \frac{ma}{K^2} \, , \quad \Omega = \frac{aK}{\alpha(r_h^2 +a^2)} -\frac{aK(1 -A^2l^2\Xi)}{l^2\Xi\alpha(1 +A^2a^2)} \, , \quad \alpha = \frac{\sqrt{(\Xi +a^2/l^2)(1 -A^2l^2\Xi)}}{1 +A^2a^2} \, ,  \\
&V = \frac{4\pi}{3\alpha K}\left\{\frac{r_h(r_h^2 +a^2)}{(1 -A^2r_h^2)^2} +\frac{m[a^2(1 -A^2l^2\Xi) +A^2l^4\Xi(\Xi +a^2l^2)]}{(1 +A^2a^2)\Xi} \right\} \, , \\
&\lambda_\pm = \frac{r_h}{\alpha(1\pm Ar_h)} +\frac{m}{\alpha\Xi^2(1 +A^2a^2)}\left[\Xi +\frac{a^2}{l^2}(2 -A^2l^2\Xi) \right]\mp \frac{Al^2(\Xi +a^2/l^2)}{\alpha(1 +A^2a^2)} \, , \\
&\mu_\pm = \frac{1}{4}\left[1 -\frac{\Xi\pm 2mA}{K} \right] \, , \qquad \Xi = 1 -\frac{a^2}{l^2} +A^2a^2 \, , \qquad P = \frac{3}{8\pi l^2} \, ,
\ea\ee
\textcolor{black}{where $V$ is the thermodynamic volume.}
It is easy to verify that the above thermodynamic quantities obey the differential first law and integral Bekenstein-Smarr mass formula simultaneously,
\bea
dM &=& TdS +\Omega dJ +VdP -\lambda_+d\mu_+ -\lambda_-d\mu_- \, , \\
M &=& 2TS +2\Omega\, J -2VP \, .
\eea

Utilizing the definition of the generalized off-shell Helmholtz free energy (\ref{FE}) and $l^2 = 3/(8\pi{}P)$, one can easily calculate the result as
\bea
\mathcal{F} &=& -\frac{(r_h^2 +a^2)\sqrt{2(8\pi P -3A^2)}(A^2a^2 +1)[(8\pi P -3A^2)r_h^2 +3]\tau}{8K\tau r_h\sqrt{\pi P}(A^2r_h^2 -1)[(3A^2 -8\pi P) +3]}  \nn \\
&&-\frac{(r_h^2 +a^2)[64\pi^2a^2Pr_h -24\pi(A^2a^2r_h +r_h)]}{8K\tau r_h(A^2r_h^2 -1)[(3A^2 -8\pi P)a^2 +3]} \, ,
\eea
Therefore, the components of the vector $\phi$ are computed as follows:
\bea
\phi^{r_h} &=& \frac{\sqrt{8\pi P -3A^2}(1 +A^2a^2)}{4K\sqrt{2\pi P}r_h^2(A^2r_h^2 -1)^2\big[(8\pi P -3A^2)a^2 -3\big]}\bigg\{8\pi Pr_h^4(A^2r_h^2 -3) -3r_h^2(A^2r_h^2 -1)^2 \nn \\
&&+a^2\Big[3(A^2r_h^2 -1)^2 -8\pi Pr_h^2(A^2r_h^2 +1)\Big]\bigg\} -\frac{2\pi A^2r_h(r_h^2 +a^2)}{K\tau(A^2r_h^2 -1)^2} +\frac{2\pi r_h}{K\tau(A^2r_h^2 -1)} \, , \\
\phi^{\Theta} &=& -\cot\Theta\csc\Theta \, .
\eea
Thus the zero point of the vector field $\phi$ is
\be
\tau = \frac{8\pi^\frac{3}{2}r_h^3\sqrt{2P(1 +A^2a^2)}\big[(3A^2 -8\pi P)a^2 +3\big]\big[(8\pi P -3A^2)(1 +A^2a^2)\big]^{-\frac{1}{2}}}{2A^2r_h^2(a^2 -r_h^2)(4\pi Pr_h^2 +3) -3A^4r_h^4(a^2 -r_h^2) +24\pi Pr_h^4 +8\pi Pa^2r_h^2 +3r_h^2 -3a^2} \, ,
\ee
which consistently reduces to the one obtained in the four-dimensional Kerr-AdS black hole case \cite{PRD107-084002} when the acceleration parameter $A$ is turned off.

\begin{figure}[t]
\subfigure[]
{\label{fig4}
\includegraphics[width=0.4\textwidth]{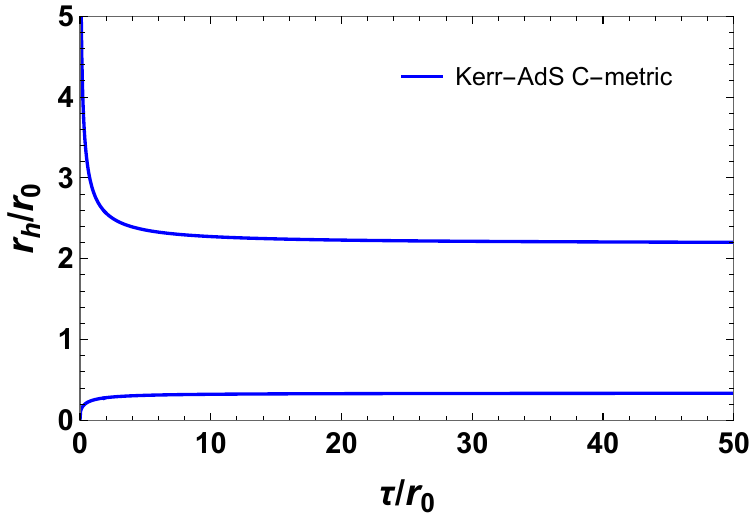}}
\subfigure[]
{\label{fig5}
\includegraphics[width=0.4\textwidth]{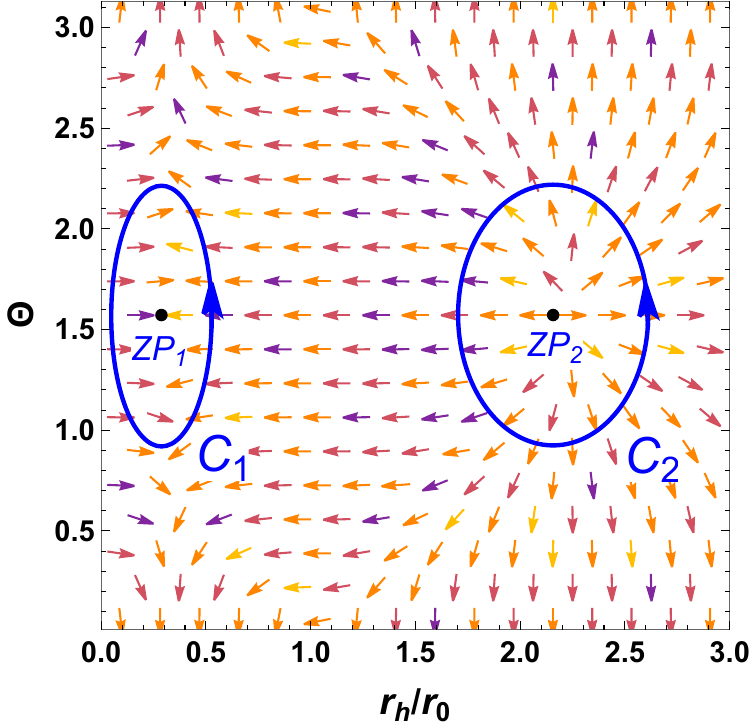}}
\caption{(a) Zero points of the vector $\phi^{r_h}$ shown in the $r_h-\tau$ plane with $a/r_0 = 1$, $Ar_0 = 1$ and $Pr_0^2 = 0.5$. \textcolor{black}{There is one thermodynamically stable and one thermodynamically unstable four-dimensional Kerr-AdS C-metric black hole branch for any value of $\tau$. Obviously, the topological number is: $W = 0$; (b) The arrows represent the unit vector field $n$ on a portion of the $r_h-\Theta$ plane for the four-dimensional Kerr-AdS C-metric black hole with $a/r_0 = 1$, $Ar_0 = 1$, $Pr_0^2 = 0.5$ and $\tau/r_0 = 20$.} The zero points (ZPs) marked with black dots are at $(r_h/r_0, \Theta) = (0.33,\pi/2)$, $(2.23,\pi/2)$, for ZP$_1$ and ZP$_2$, respectively. The blue contours $C_i$ are closed loops surrounding the zero points.}
\end{figure}

For the Kerr-AdS C-metric black hole, the zero points of the component $\phi^{r_h}$ can be plotted with $a/r_0 = 1$, $Ar_0 = 1$, and $Pr_0^2 = 0.5$ in Fig. \ref{fig4}, while the unit vector field $n$ is depicted on a section of the $\Theta-r_h$ plane in Fig. \ref{fig5} with $\tau/r_0 = 20$.
As illustrated in Fig. \ref{fig4}, for any given value of $\tau$, there consistently exist two Kerr-AdS C-metric black holes: one thermodynamically stable and the other thermodynamically unstable.
In Fig. \ref{fig5}, two zero points are located at $(r_h/r_0, \Theta) = (0.33,\pi/2)$ and $(2.23,\pi/2)$, respectively.
The winding numbers $w_i$ for the blue contours $C_i$ are found to be $w_1 = -1$ and $w_2 = 1$, in contrast to the four-dimensional Kerr-AdS black hole, which only has $w_1 = 1$.
Consequently, the topological number $W = 0$ for the Kerr-AdS C-metric black hole, as evident in Fig. \ref{fig5}, differs from the topological number of the four-dimensional Kerr-AdS black hole ($W = 1$) \cite{PRD107-084002}.
This suggests that the topological number is significantly influenced by the acceleration parameter.

Furthermore, in light of another intriguing finding presented in Ref.  \cite{PRD108-084041}, which demonstrate that the topological number of an non-rotating, accelerating AdS black hole is consistently lower by $-1$ compared to its non-accelerating counterpart, a broader conjecture within the framework of general relativity is proposed: the topological number of an accelerating AdS black hole consistently differs by minus one from that of its corresponding non-accelerating AdS black hole.
This conjecture will be further explored in section \ref{4.2}, where we calculate the topological number of the KN-AdS C-metric black hole and compare it to that of the KN-AdS black hole, thereby affirming the validity and relevance of our proposal.

In addition, since the Kerr-AdS C-metric black hole shares the same topological number as the Kerr black hole ($W = 0$), we demonstrate that although the acceleration parameter and the negative cosmological constant each independently raise the topological number by one, their combined presence in the rotating black hole solution neutralizes their effects.

\section{Topological classes of rotating charged accelerating black hole}\label{IV}
In this section, we turn to explore the topological number of the four-dimensional rotating charged accelerating black hole by considering the KN C-metric solution, and then extend it to the KN-AdS C-metric case with a nonzero negative cosmological constant.

\subsection{KN C-metric black hole}\label{4.1}
The metric and the Abelian gauge potential of the KN C-metric black hole are still given by Eqs. (\ref{KNAdSCmetric}) and (\ref{Abelian}), however, the metric functions are respectively
\be\ba
f(r) &= (1 -A^2r^2)(r^2 -2mr +a^2 +q^2) \, ,   \\
h(\theta) &= 1 +2mA\cos\theta +A^2(a^2 +q^2)\cos^2\theta \, .
\ea\ee

The thermodynamic quantities are given by \cite{JHEP0419096}
\be\ba\label{ThermKNC}
&M = \frac{m(1 -A^2a^2)}{\alpha K(1 +A^2a^2)} \, , ~ \qquad T = \frac{\p_{r_h}f(r_h)}{4\pi\alpha(r_h^2 +a^2)} \, , ~ \qquad S = \frac{\pi(r_h^2 +a^2)}{K(1 -A^2r_h^2)} \, ,    \\
&\alpha = \frac{\sqrt{(1 -A^2a^2)\Xi}}{1 +A^2a^2} \, , ~ \qquad J = \frac{ma}{K^2} \, , ~ \qquad \mu_{\pm} = \frac{1}{4}\left[1 -\frac{\Xi\pm 2mA}{K} \right] \, ,  \\
&\Omega = \frac{aK}{\alpha(r_h^2 +a^2)} -\frac{A^2aK}{\alpha(1 +A^2a^2)} \, , \quad \lambda_\pm = \frac{r_h}{\alpha(1\pm Ar_h)} -\frac{MK}{1 +A^2a^2} \mp\frac{Aa^2}{\alpha(1 +A^2a^2)} \, , \quad \\
&Q = \frac{q}{K} \, , ~\qquad \Phi = \frac{qr_h}{\alpha(r_h^2 +a^2)} \, , ~\qquad \Xi = 1 +A^2(a^2 +q^2) \, ,
\ea\ee
\textcolor{black}{where $Q$ and $\Phi$ denote the electric charge and its corresponding electrostatic potential at the event horizon, respectively.}
Nevertheless, we can verify that both the differential and integral mass formulas are completely satisfied
\bea
dM &=& TdS +\Omega dJ +\Phi dQ -\lambda_+d\mu_+ -\lambda_-d\mu_- \, , \\
M &=& 2TS +2\Omega\, J +\Phi Q \, .
\eea

Next, we will discuss the topological number of the KN C-metric black hole. It is simple to derive the generalized off-shell Helmholtz free energy as
\be
\mathcal{F} = M -\frac{S}{\tau} = \frac{(r_h^2 +a^2 +q^2)(1 -A^2a^2)}{2\alpha Kr_h(1 +A^2a^2)} -\frac{\pi(r_h^2 +a^2)}{K\tau(1 -A^2r_h^2)} \, .
\ee
Then, the components of the vector $\phi$ are
\bea
\phi^{r_h} &=& \frac{(r_h^2 -a^2 -q^2)(1 -A^2a^2)}{2\alpha Kr_h^2(1 +A^2a^2)} -\frac{2\pi r_h(1 +A^2a^2)}{K\tau(A^2r_h^2 -1)^2} \, , \\
\phi^{\Theta} &=& -\cot\Theta\csc\Theta \, .
\eea
Thus, by solving the equation: $\phi^{r_h}$ = 0, one can easily obtain
\be
\tau = \frac{4\pi\alpha r_h^3(1 +A^2a^2)^2}{(r_h^2 -a^2 -q^2)(1 -A^2a^2)(A^2r_h^2 -1)^2} \ee
as the zero point of the vector field $\phi$, which consistently reduces to the one obtained in the four-dimensional KN black hole \cite{PRD107-024024} when the acceleration parameter $A$ vanishes.

\begin{figure}[t]
\subfigure[]
{\label{fig6}
\includegraphics[width=0.4\textwidth]{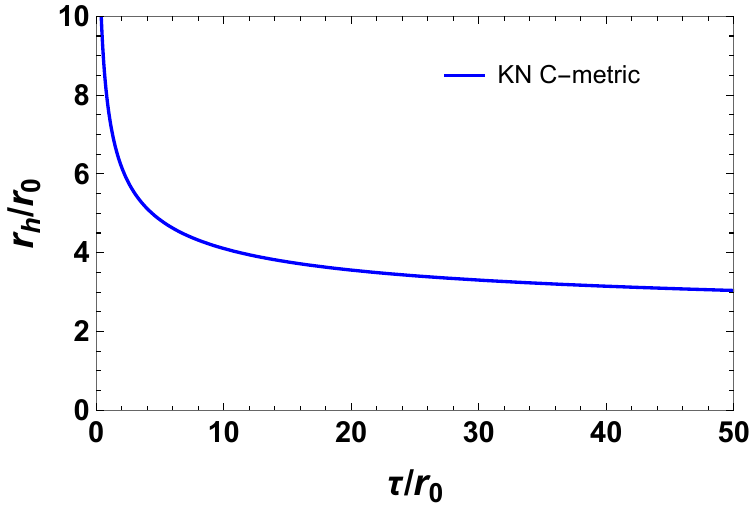}}
\subfigure[]
{\label{fig7}
\includegraphics[width=0.4\textwidth]{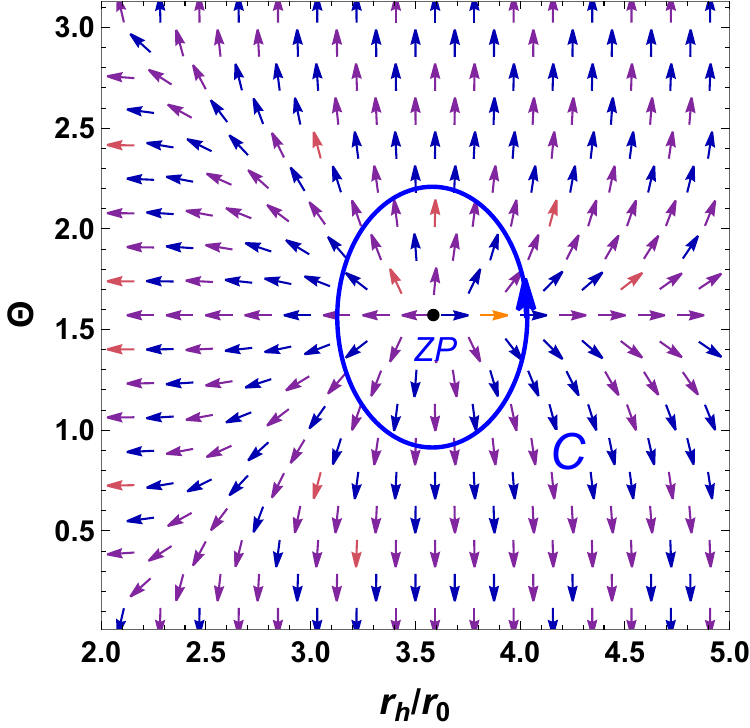}}
\caption{(a) Zero points of the vector $\phi^{r_h}$ shown in the $r_h-\tau$ plane with $a/r_0 = 1$, $q/r_0 = 1$ and $Ar_0 = 0.5$. \textcolor{black}{There is always one thermodynamically stable four-dimensional KN C-metric black hole for any value of $\tau$. Obviously, the topological number is: $W = 1$; (b) The arrows represent the unit vector field $n$ on a portion of the $r_h-\Theta$ plane for the four-dimensional KN C-metric black hole with $a/r_0 = 1$, $q/r_0 = 1$, $Ar_0 = 0.5$ and $\tau/r_0 = 20$.} The zero point (ZP) marked with black dot is at $(r_h/r_0, \Theta) = (3.56,\pi/2)$. The blue contours $C$ is closed loop enclosing the zero point.}
\end{figure}

Considering $a/r_0 = 1$, $q/r_0 = 1$ and $Ar_0 = 0.5$ for the KN C-metric black hole, we plot the zero points of $\phi^{r_h}$ in the $r_h-\tau$ plane in Fig. \ref{fig6}, and the unit vector field $n$ on a portion of the $\Theta-r_h$ plane with $\tau/r_0 = 20$ in Fig. \ref{fig7}.
Evidently, for any value of $\tau$, there exists a unique thermodynamically stable KN C-metric black hole.
As shown in Fig. \ref{fig7}, the zero point is located at $(r_h/r_0, \Theta) = (3.56,\pi/2)$.
By analyzing the local properties of this zero point, we can readily determine the topological number $W = 1$ for the KN C-metric black hole.
Comparing this result with the corresponding finding for the Kerr C-metric black hole in section \ref{3.1}, where $W = 1$ as well, it becomes evident that the presence of the electric charge parameter does not influence the topological number of rotating, accelerating black holes.
This observation further supports our initial conjecture, as proposed in Ref. \cite{PRD107-024024}, that the topological number remains unaffected by the charge in rotating black holes--a conclusion that not only extends to the case of rotating black holes in gauged supergravity theories \cite{PLB856-138919}, but also to the case of rotating, accelerating black holes.
In addition, by comparing with the corresponding result for the KN black hole ($W = 0$), we further validate the conjecture proposed in section \ref{3.1}: the difference in topological numbers between any asymptotically flat accelerating black hole and its non-accelerating counterpart consistently equals one.
This conclusion also holds true in the case of rotating, accelerating charged black holes.

\subsection{KN-AdS C-metric black hole}\label{4.2}
In this subsection, we consider the general KN-AdS C-metric black hole case.
The metric, the Abelian gauge potential, and the metric functions are already given in Eqs. (\ref{KNAdSCmetric})-(\ref{mf}).
The thermodynamic quantities of the KN-AdS C-metric black hole are \cite{JHEP0419096}
\be\ba\label{ThermKNAdSC}
&M = \frac{m(\Xi +a^2/l^2)(1 -A^2l^2\Xi)}{\Xi\alpha K(1 +A^2a^2)} \, , \quad T = \frac{\p_{r_h}f(r_h)}{4\pi\alpha(r_h^2 +a^2)} \, , ~ \quad S = \frac{\pi(r_h^2 +a^2)}{K(1 -A^2r_h^2)} \, , \quad Q = \frac{q}{K} \, ,   \\
&J = \frac{ma}{K^2} \, , \quad \Omega = \frac{aK}{\alpha(r_h^2 +a^2)} -\frac{aK(1 -A^2l^2\Xi)}{l^2\Xi\alpha(1 +A^2a^2)} \, , \quad \alpha = \frac{\sqrt{(\Xi +a^2/l^2)(1 -A^2l^2\Xi)}}{1 +A^2a^2} \, ,  \\
&V = \frac{4\pi}{3\alpha K}\left\{\frac{r_h(r_h^2 +a^2)}{(1 -A^2r_h^2)^2} +\frac{m[a^2(1 -A^2l^2\Xi) +A^2l^4\Xi(\Xi +a^2l^2)]}{(1 +A^2a^2)\Xi} \right\} \, , \quad \Phi = \frac{qr_h}{\alpha(r_h^2 +a^2)} \, , \\
&\lambda_\pm = \frac{r_h}{\alpha(1\pm Ar_h)} +\frac{m}{\alpha\Xi^2(1 +A^2a^2)}\left[\Xi +\frac{a^2}{l^2}(2 -A^2l^2\Xi) \right]\mp \frac{Al^2(\Xi +a^2/l^2)}{\alpha(1 +A^2a^2)} \, , \\
&\mu_\pm = \frac{1}{4}\left[1 -\frac{\Xi\pm 2mA}{K} \right] \, , \qquad \Xi = 1 -\frac{a^2}{l^2} +A^2(a^2 +q^2) \, , \qquad P = \frac{3}{8\pi l^2} \, , \quad
\ea\ee
It is simple to prove that the above thermodynamic quantities (\ref{ThermKNAdSC}) obey the first law and the Bekenstein-Smarr mass formula simultaneously
\bea
dM &=& TdS +\Omega dJ +\Phi dQ +VdP -\lambda_+d\mu_+ -\lambda_-d\mu_- \, , \\
M &=& 2TS +2\Omega\, J +\Phi Q -2VP \, .
\eea

\begin{figure}[t]
\subfigure[]
{\label{fig8}
\includegraphics[width=0.4\textwidth]{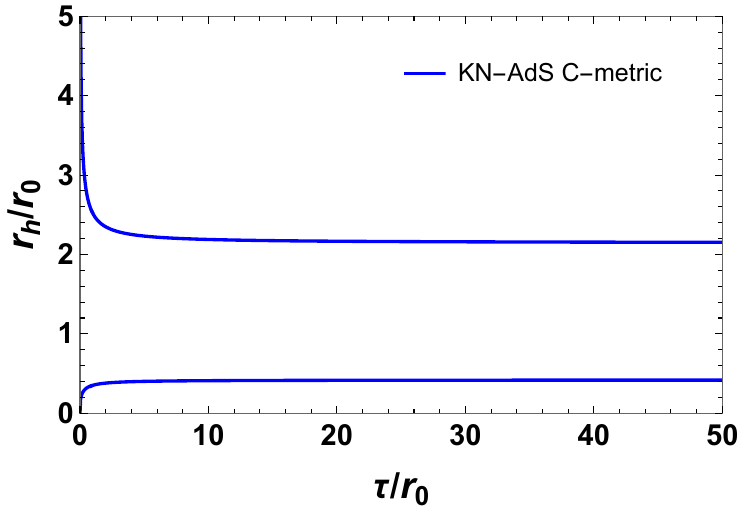}}
\subfigure[]
{\label{fig9}
\includegraphics[width=0.4\textwidth]{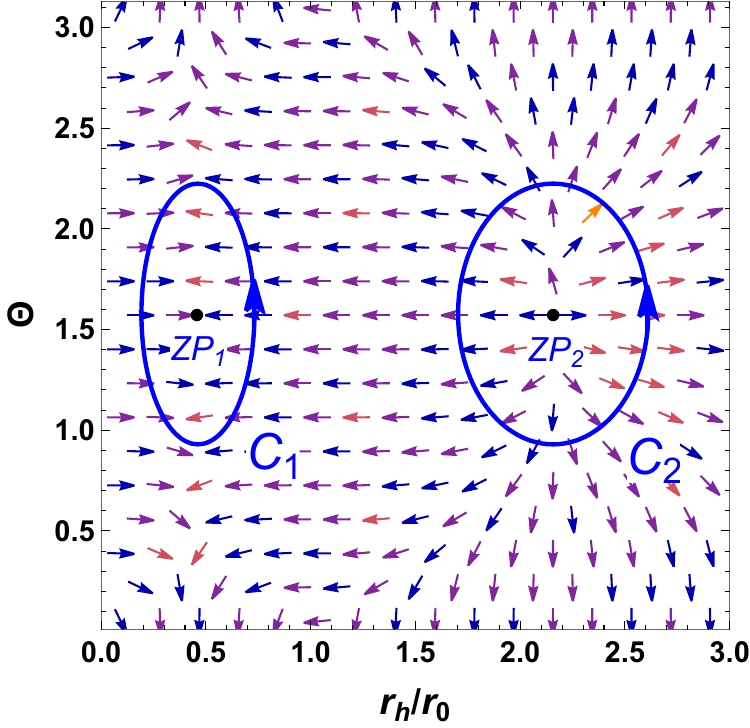}}
\caption{(a) Zero points of the vector $\phi^{r_h}$ shown in the $r_h-\tau$ plane with $a/r_0 = 1$, $q/r_0 = 1$, $Ar_0 = 1$ and $Pr_0^2 = 0.5$. \textcolor{black}{There is one thermodynamically stable and one thermodynamically unstable four-dimensional KN-AdS C-metric black hole branch for any value of $\tau$. Obviously, the topological number is: $W = 0$; (b) The arrows represent the unit vector field $n$ on a portion of the $r_h-\Theta$ plane for the four-dimensional KN-AdS C-metric black hole with $a/r_0 = 1$, $q/r_0 = 1$, $Ar_0 = 1$, $Pr_0^2 = 0.5$ and $\tau/r_0 = 20$.} The zero points (ZPs) marked with black dots are at $(r_h/r_0, \Theta) = (0.42,\pi/2)$, $(2.17,\pi/2)$, for ZP$_1$ and ZP$_2$, respectively.
The blue contours $C_i$ are closed loops surrounding the zero points.}
\end{figure}

In the following, we will investigate the topological number of the KN-AdS C-metric black hole.
The generalized off-shell Helmholtz free energy is given by
\bea
\mathcal{F} &=& \frac{\sqrt{1 +A^2(a^2 +q^2)}}{K[1 -8\pi Pa^2/3 +A^2(a^2 +q^2)]}\sqrt{1 +\frac{A^2[-3 +8\pi Pa^2 -3A^2(a^2 +q^2)]}{8\pi P}}\bigg[\frac{r_h^2 +a^2 +q^2}{2r_h} \nn \\
&&+\frac{4\pi Pr_h(r_h^2 +a^2)}{3 -3A^2r_h^2} \bigg] +\frac{\pi(r_h^2 +a^2)}{K\tau(A^2r_h^2 -1)} \, . \quad
\eea
Therefore the components of the vector $\phi$ can be derived as
\bea
\phi^{r_h} &=& \frac{\sqrt{1 +A^2(a^2 +q^2)}}{6K[1 -8\pi Pa^2/3 +A^2(a^2 +q^2)]}\sqrt{1 +\frac{A^2[-3 +8\pi Pa^2 -3A^2(a^2 +q^2)]}{8\pi P}}\Bigg\{3 -\frac{3q^2}{r_h^2} -\frac{8\pi Pr_h^2(A^2r_h^2 -3)}{(A^2r_h^2 -1)^2} \nn \\
&& +a^2\Bigg[4\pi P\bigg(\frac{1}{(Ar_h -1)^2} +\frac{1}{(Ar_h +1)^2} \bigg) \Bigg] \Bigg\} -\frac{2\pi A^2r_h(r_h^2 +a^2)}{K\tau(A^2r_h^2 -1)^2} -\frac{2\pi r_h}{K\tau(1 -A^2r_h^2)} \, , \nn \\
\phi^{\Theta} &=& -\cot\Theta\csc\Theta \, .
\eea
Solving the equation $\phi^{r_h} = 0$ straightforwardly yields
\bea
\tau &=& -\frac{8\sqrt{2P}\pi^{\frac{3}{2}}r_h^3(1 +A^2a^2)\big[3 -8\pi Pa^2 +3A^2(a^2 +q^2) \big]}{\sqrt{\big[1 +A^2(a^2 +q^2)\big]\big[8\pi P +A^2(8\pi Pa^2 -3) -3A^4(a^2 +q^2)\big]}}\Big\{-3(r_h^2 +8\pi Pr_h^4) +3q^2(A^2r_h^2 -1)^2  \nn \\
&&+A^2r_h^4\big[r_h^2(8\pi P -3A^2) +6\big] +a^2\big[3(A^2r_h^2 -1)^2 -8\pi Pr_h^2(1 +A^2r_h^2)\big]\Big\}^{-1}
\eea
as the zero point of the vector field $\phi$, consistent with the result for the four-dimensional KN-AdS black hole when the acceleration parameter $A$ vanishes, as detailed in Ref. \cite{PRD107-084002}.

In Figs. \ref{fig8} and \ref{fig9}, taking $a/r_0 = 1$, $q/r_0 = 1$, $Ar_0 = 1$, $Pr_0^2 = 0.5$ for the KN-AdS C-metric black hole, we plot the zero points of $\phi^{r_h}$ in the $r_h-\tau$ plane and the unit vector field $n$ with $\tau = 20r_0$, respectively.
Note that for these values of $a/r_0$, $q/r_0$, $Ar_0$ and $Pr_0^2$, there are one thermodynamically stable and one thermodynamically unstable KN-AdS C-metric black hole for any value of $\tau$.
In Fig. \ref{fig9}, one can observe that two zero points are located at $(r_h/r_0, \Theta) = (0.42,\pi/2)$ and $(2.17,\pi/2)$, respectively.
As a result, the topological number $W = 0$ for the KN-AdS C-metric black hole can be explicitly established in Figs. \ref{fig8} and \ref{fig9} via the local property of the zero point, which is different from that of the four-dimensional KN-AdS black hole ($W = 1$) \cite{PRD107-084002}.
Therefore, we further verify that the conjecture proposed in section \ref{3.2}--"the topological number of an asymptotically AdS, accelerating black hole consistently differs by minus one from that of its corresponding asymptotically AdS, non-accelerating black hole"--is also applicable to the case of rotating, accelerating charged AdS black holes.
Furthermore, since the KN-AdS C-metric black hole retains the same topological number as the KN black hole ($W = 0$), we show that although both the acceleration parameter and the negative cosmological constant individually increase the topological number by one, their combined effect in the rotating charged black hole solution counteracts this increase.

\section{Concluding remarks}\label{V}
\begin{table}[t]
\centering
\caption{The topological number $W$, numbers of generation and annihilation points for accelerating black holes and their usual nonaccelerating counterparts.}
\resizebox{0.6\textwidth}{!}{
\begin{tabular}{c|c|c|c}
\hline\hline
BH solution & $W$ & Generation point & Annihilation point\\ \hline
Schwarzschild \cite{PRL129-191101} & -1 & 0 & 0 \\
Schwarzschild-AdS \cite{PRD106-064059} & 0 & 0 & 1 \\
C-metric \cite{PRD108-084041} & 0 & 0 & 0\\
AdS-C-metric \cite{PRD108-084041} & -1 & 1 or 0 & 1 or 0\\ \hline
RN \cite{PRL129-191101} & 0 & 1 & 0 \\
RN-AdS \cite{PRL129-191101} & 1 & 1 or 0 & 1 or 0 \\
RN-C-metric \cite{PRD108-084041} & 1 & 0 & 0\\
RN-AdS-C-metric \cite{PRD108-084041} & 0 & 1 & 0 \\ \hline
Kerr \cite{PRD107-024024} & 0 & 1 & 0 \\
Kerr-AdS \cite{PRD107-084002} & 1 & 1 or 0 & 1 or 0 \\
Kerr C-metric & 1 & 0 & 0 \\
Kerr-AdS C-metric & 0 & 0 & 0 \\ \hline
KN \cite{PRD107-024024} & 0 & 1 & 0 \\
KN-AdS \cite{PRD107-084002} & 1 & 0 & 0 \\
KN C-metric & 1 & 0 & 0 \\
KN-AdS C-metric & 0 & 0 & 0  \\
\hline\hline
\end{tabular}}
\label{TableI}
\end{table}

Our results found in the present paper are now summarized in the following Table \ref{TableI}.

In this paper, by employing the generalized off-shell Helmholtz free energy, we explore the topological characteristics of rotating, accelerating black holes.
Specifically, we analyze the topological number of the Kerr C-metric and Kerr-AdS C-metric black holes.
Additionally, we apply the same approach to examine the topological number of the KN C-metric and KN-AdS C-metric black holes.
Our findings reveal that the Kerr C-metric and KN C-metric black holes share the same topological classification, both having a topological number of $W = 1$, while the Kerr-AdS C-metric and KN-AdS C-metric black holes belong to a different topological class, characterized by a topological number of $W = 0$.

Building on the results presented in this paper and our previous study \cite{PRD108-084041}, we identify three intriguing outcomes: I) The topological number of an asymptotically flat accelerating black hole consistently differs by one from that of its non-accelerating counterpart.
II) For an asymptotically AdS accelerating black hole, the topological number is reduced by one compared to the corresponding non-accelerating AdS black hole.
III) In the context of general relativity, the acceleration parameter and the negative cosmological constant each independently raise the topological number by one.
However, when both are present, their influences counteract, leaving the topological number unchanged.

\textcolor{black}{In summary, by studying the thermodynamic topological classification of rotating and accelerating black holes, we find that their thermodynamic phase transition behaviors can be uniformly characterized through topological invariants (e.g., winding numbers, topological numbers), revealing the regulatory mechanisms of angular momentum and acceleration on their thermodynamic properties. By transforming phase transition problems into the analysis of structures in topological space, this classification not only systematizes our understanding of known phase transitions (such as the Hawking-Page transition and P-V criticality) but also predicts novel thermodynamic phase transition behaviors \cite{2402.00106,2405.07525}, bridging the connection between macroscopic thermodynamics and microscopic quantum structures. This approach is increasingly becoming a crucial tool for exploring the nature of black holes and gravity theories.}

There are two promising directions for future research emerge from this work.
The first is to investigate the topological numbers of black holes in alternative theories of gravity, including scalar-tensor-vector gravity \cite{Moffat:2005si,Liu:2023uft,2406.00579,Qiao:2024gfb}, bumblebee gravity \cite{Casana:2017jkc,Liu:2022dcn,Liu:2024oeq}, and Kalb-Ramond gravity \cite{Yang:2023wtu,2406.13461,2407.07416} (\textcolor{black}{additional fields may introduce new topological defects, potentially altering the value of $W$}). We have noted that Refs. \cite{2409.09333,2409.12747,2411.10102,2504.10858} recently proposed a universal thermodynamic topological classification method for black hole solutions. Therefore, the second outlook is to explore the universal thermodynamic topological classes of rotating and accelerating black holes.

\acknowledgments
\textcolor{black}{We are greatly indebted to the anonymous referees for their constructive comments to improve the presentation of this work.} This work is supported by the National Natural Science Foundation of China (NSFC) under Grants No. 12205243, No. 12375053, No. 12475051, No. 12122504, and No. 12035005; the Sichuan Science and Technology Program under Grant No. 2023NSFSC1347; the Doctoral Research Initiation Project of China West Normal University under Grant No. 21E028; the science and technology innovation Program of Hunan Province under Grant No. 2024RC1050; the innovative research group of Hunan Province under Grant No. 2024JJ1006; and the Hunan Provincial Major Sci-Tech Program under Grant No.2023ZJ1010.

\end{document}